\documentstyle[12pt,graphicx]{article}
\title{ 3-loop Yang-Mills Condensate Dark Energy Model
               And Its Cosmological Constraints}
\author{S. Wang\footnote{Email: swang@mail.ustc.edu.cn}
        \, Y. Zhang and T.Y. Xia \\
        Astrophysics Center \\
        University of Science and Technology of China \\
        Hefei, Anhui, China }
 \date{}

\begin{document}
\maketitle
\baselineskip=21truept

\newcommand{\be}{\begin{equation}}
\newcommand{\ee}{\end{equation}}


\sf
\begin{center}
\Large  Abstract
\end{center}
\begin{quote}
 {\large

This work is a comprehensive investigation of the Yang-Mills condensate (YMC) dark energy (DE) model,
which is extended to include the 3-loop quantum corrections.
We study its cosmic evolution and the possibility of crossing phantom divide $w=-1$,
examine in details the Hubble parameter $H$, the deceleration parameter $q$,
the statefinder diagnosis $(r,s)$, and the $w-w^\prime$ diagnosis of the model without and with interaction,
and compare our results with other DE models.
Besides, by using the observational data of type Ia supernovae (SNIa),
the shift parameter from cosmic microwave background (CMB),
and the baryon acoustic oscillation (BAO) peak from large scale structures (LSS),
we give the cosmological constraints on 3-loop YMC model.
It is found that the model can naturally solve the coincidence problem,
and its prediction of the afore-mentioned parameter
is much closer to the $\Lambda$CDM model than other dynamics DE models;
the introduction of the matter-DE interaction will make the YMC model deviating from the $\Lambda$CDM model,
and will give an equation of state (EOF) crossing $-1$.
Moreover, it is also found that,
to fit the latest SNIa data alone,
the $\Lambda$CDM model is slightly better than the 3-loop YMC model;
but in fitting of the combination of SNIa, CMB and LSS data,
the 3-loop YMC model performs better than the $\Lambda$CDM model.}

\end{quote}


\newpage
\baselineskip=21truept

\begin{center}
{\em\Large 1. Introduction}
\end{center}

Observations on the type Ia Supernova \cite{Riess1},
CMB anisotropies \cite{spergel} and large scale structure \cite{bahcall}
all indicate the existence of mysterious dark energy
that driving the current accelerating expansion of Universe.
To interpret the physics of dark energy,
there have been a large amount of models proposed.
The simplest one is the $\Lambda$CDM model,
which can fit the observations so far,
but is plagued with the fine-tuning problem and the coincidence problem \cite{Weinberg}.
While the former problem exists for almost all the DE models,
the latter one can be solved in the frame of dynamical DE models.
Among them are
quintessence \cite{quint}, phantom \cite{phantom}, k-essence \cite{k}, quintom\cite{quintom},
tachyonic \cite{tachyonic}, holographic \cite{holographic}, agegraphic \cite{agegraphic}, ect.
There are also other interesting models either based on the effective gravity \cite{raval},
or on the Born-Infeld quantum condensate \cite{Elizalde}.
In our previous works \cite{zhang0} a dynamic model is proposed,
in which the renormalization-group improved effective YMC serves as the dark energy;
and in a latest work \cite{Xia}, this model was extend to the 2-loop quantum corrections.
Unlike the scalar models,
our model is based on a vector-type quantum effective Yang-Mills (YM) fields
and does not suffer from the difficulties of scalar models mentioned in Ref.\cite{padmanabhan}.
The effective YMC in our model is
a coherent boson field system at low temperatures,
and the DE is viewed as the ground state energy of this YMC field.
As is known, a quantum system of bosons with a conservative ``charge''
(such as the particle number, the electric charge, the color in QCD, etc)
will experience Bose-Einstein condensation when
the temperature is low enough or the charge density is high enough.
This applies to many systems, such as Q-balls \cite{Lee,Coleman},
charged relativistic scalar bosons \cite{Weldon,Parker},
and the gluons condensate in the effective QCD models \cite{pagels,adler}.
And this is the physical origin of our Yang-Mills condensate model.
From the viewpoint of quantum effective field theory at low temperatures,
it would be desired if one can include high order quantum corrections
as much as possible.
In this work we will extend the YMC DE model to the 3-loop quantum corrections,
focusing on its cosmic evolution and the issue of crossing $w = -1$.

As an important next step,
one needs to confront DE models with observational data.
In a recent work, by using the differential ages of passively evolving galaxies,
Simon et al. gave 9 observational $H(z)$ data points in the range $0\leq z \leq 1.8$ \cite{Simon},
which have been used to constrain various DE models \cite{Samushia,Hao1}.
Besides, the deceleration parameter $q(z)$,
containing the second order derivative of the scale factor $a(t)$,
is also important in confronting DE models \cite{John,Virey,Gong}.
Moreover, a new geometrical diagnosis pair $(r,s)$,
called statefinder,
has been introduced to distinguish DE models \cite{Sahni}.
Since the pair $(r,s)$ contains the third order derivative of $a(t)$
and their values is expected to be available
from the future SNAP observation \cite{Alam},
the statefinder diagnosis has attracted a lot of attention
\cite{Gorini,Zimdahl,XZ,Setare,Wu,Chang,Yi,Hu,Zhao1,Hao2,JFZ}.
Finally, another dynamical diagnosis $w-w'$,
consisting of the EOS and its time derivative,
are also extensively used in the literatures
\cite{Caldwell,Linder,Scherrer,Chiba,Barger,Zhao2,Calcagni,Guo,Huang,Putter}.
In viewing these different diagnosis methods,
we shall present a comprehensive analysis for
$H(z)$, $q(z)$, $(r,s)$ and $w-w'$, respectively, in the 3-loop YMC model.
Comparing with the issues listed above,
it is more common to test various DE models by using the observational data of SNIa.
Especially, in a recent work,
Riess et al released the up-to-data 182 gold sample of SNIa \cite{Riess2},
which has been used in Refs. \cite{XZ2,XW,Hao3}.
In addition, as useful complements to that of SNIa,
the shift parameter $R$ from CMB observations \cite{Tegmark,ywang}
and the BAO peak parameter $A$ from LSS \cite{Eisenstein}
are also crucial to constrain various DE models.
So in this work,
we shall utilize the combination of SNIa, CMB, and LSS data,
to give the cosmological constraints on our model.

The organization of this paper is as follows.
In section 2, we extend our previously proposed YMC DE model to the 3-loop quantum corrections,
present its prediction of cosmic evolution,
and explore the possibility of crossing the phantom divide $w = -1$ .
In section 3, we study the Hubble parameter $H$, the deceleration parameter $q$,
the statefinder diagnosis $(r,s)$, and the $w-w^\prime$ diagnosis in 3-loop YMC model,
and compare our results with other DE models.
In Section 4, the observational data of SNIa, CMB, and LSS are employed
to give the cosmological constraints on the 3-loop YMC model.
Section 5 is a short summary.
In this paper the unit with $c=\hbar=1$ is used.

\

\begin{center}
{\em\Large 2. 3-loop YMC Model}
\end{center}

In the renormalization-group improved effective YM field theory \cite{savvidi,adler2},
the running coupling constant up to the 3-loop \cite{Politzer,Weinberg2},
should have the following form
\be
\label{g} g^{2}(F)= \frac{1}{b}\left[\frac{1}{\tau}
-\eta\frac{\ln|\tau|}{\tau^{2}}
+\eta^{2}\frac{\ln^2|\tau|-\ln|\tau|+C}{\tau^{3}}
+O(\frac{1}{\tau^{3}})\right],
\ee
where $\tau\equiv\ln|F/e\kappa^2|$, $F\equiv -\frac{1}{2} F^a{}_{\mu\nu}F^{a}{}^{\mu\nu}= E^{2}-B^{2}$
plays the role of the order parameter of the YMC,
and the parameter $\kappa$ is the renormalization scale with dimension of squared mass.
For the gauge group $SU(N)$ without fermions,
$b=\frac{11N}{3(4\pi)^2}$,
$\eta \equiv \frac{2\beta_{1}}{\beta_{0}^{2}}\simeq 0.8$.
Here $C\equiv\frac{8\beta_{0}\beta_{2}}{\beta_{1}^{2}}-1$,
and the numerical coefficients $\beta_{0}$, $\beta_{1}$, $\beta_{2}$ are given in Ref.\cite{Weinberg2}.
It should be stressed that, although the 1-loop and 2-loop corrections are uniquely fixed,
the 3-loop correction to $g^2$ is renormalization-scheme dependent,
so is the coefficient $C$ \cite{muta}.
Notice that the Lorentz invariance is reserved in the effective YM theory,
because the Lagrangian is constructed out of the combinations
of $F^a\,_{\mu\nu} F^a\,^{\mu\nu}$ \cite{pagels,adler,Zhang}.
For simplicity, we only discuss the case of pure "electric" condensate with $F=E^{2}$
(the case of including magnetic component was discussed in Ref. \cite{Zhao3}).
The effective Lagrangian, defined as $L_{eff}=F/2g^2(F)$, is given by
\be \label{Lagrangian}
L_{eff}=\frac{1}{2}b \kappa^2 e^{y}
\left[(y-1)+\eta\ln|y-1+\delta|-\eta^{2}\frac{\ln^2|y-1
+\delta|-\ln|y-1+\delta|}{y-1+\delta}\right],
\ee
where the variable $y\equiv \tau+1=\ln|F/\kappa^2|$,
and $\delta$, a dimensionless parameter, represents the higher order corrections,
including the type of terms, such as the term of $C/\tau^{3}$ in Eq.(\ref{g}).
In the bracket of Eq.(\ref{Lagrangian}) the $\eta$ term is the 2-loop contribution,
and the $\eta^{2}$ term is the 3-loop one,
which adds no new parameter other than the 2-loop model.
Notice that the YM field introduced here is not the gluon fields in QCD,
nor the gauge boson fields in the electro-weak unification.
From the effective Lagrangian in Eq.(\ref{Lagrangian}),
follow the energy density and the pressure of YMC DE
\be \label{energy}
\rho_y=\frac{1}{2}b \kappa^2e^{y}\left[(y+1)+\eta
(Y_{1}+2Y_{2})-\eta^{2}(Y_{3}-2Y_{4})\right],
\ee
\be
\label{pressure} p_y=\frac{1}{6}b \kappa^2e^{y}\left[(y-3)+\eta
(Y_{1}-2Y_{2})-\eta^{2}(Y_{3}+2Y_{4})\right],
\ee
where
\be \label{y_{1}}
Y_{1} \equiv\ln|y-1+\delta|, \,\,\,\,\,
Y_{2}\equiv\frac{1}{y-1+\delta},
\ee
\be
\label{y_{3}}
Y_{3}\equiv (Y_1-1)Y_1Y_2, \,\,\,\,
Y_{4}\equiv (Y_1-3)Y_1Y_2^2.
\ee
As a consistency check, the well-known conformal trace anomaly \cite{pagels,Collins}
\be T^\mu \, _\mu
=\rho_y-3p_y
  = 2F\frac{d}{d\tau} \left[\frac{1}{g^2(F)} \right]
\ee
is satisfied up to the 3-loop.
It is known that the trace anomaly occurs as a quantum effect of
the vacuum and only violates the traceless condition
of the stress tensor $T_{\mu\nu}$ without violating the Lorentz invariance.
Also the form of the stress tensor $T_{\mu\nu}$ of YM fields
is consistent with homogeneity and isotropy of the Universe.
The EOS for the YMC is given by
\be
\label{w} w=\frac{p_y}{\rho_y} \, .
\ee
If one ignores the terms of $\eta^{2}$ from Eq.(\ref{g}) through Eq.(\ref{pressure}),
the 2-loop model \cite{Xia} is obtained,
and if one further sets $\eta=0$, the 1-loop model \cite{zhang0} is recovered.

In our model the Universe is filled with three kinds of major energy components:
the dark energy represented by the YMC,
the matter (baryons and dark matter),
and the radiation (consisting of CMB and other massless particles).
The overall cosmic expansion is determined by the Friedmann equation
\be \label{friedmann}
(\frac{\dot{a}}{a})^2=\frac{8 \pi G}{3}(\rho_y+\rho_m+\rho_r),
\ee
where $\rho_m$ is the energy density of the matter,
and $\rho_r$ is of the radiation.
The dynamical evolutions of the three components are given by
\be \label{ymeq}
\dot{\rho}_y+3\frac{\dot{a}}{a}(\rho_y+p_y)=-\Gamma \rho_y,
\ee
\be
\label{meq} \dot{\rho}_m+3\frac{\dot{a}}{a}\rho_m=\Gamma \rho_y,
\ee
\be \label{req} \dot{\rho}_r+3\frac{\dot{a}}{a}(\rho_r+p_r)=0,
\ee
where $p_r$ is the radiation pressure,
$\Gamma $ is the decay rate of the YMC into matter, a parameter of the model.
If $\Gamma=0$, the YMC does not couple to the matter;
if $\Gamma>0$, the interaction term $\Gamma\rho_y$ in Eqs.(\ref{ymeq}) and (\ref{meq})
represents the rate of energy transfer from the YMC to the matter.
The sum of Eqs.({\ref{ymeq}}), ({\ref{meq}}), and ({\ref{req}})
guarantees  that the total energy of the three components is still conserved.
Replacing the old variables $t$, ${\rho}_m$, ${\rho}_r$, and ${\rho}_y$
with new variables $N \equiv\ln a(t)$, $x \equiv \rho_m/ \frac{1}{2}b \kappa^2$,
$r\equiv \rho_r/ \frac{1}{2}b \kappa^2$, and $y$,
and making use of the Friedmann equation at $z=0$,
the set of equations, (\ref{ymeq}) through ({\ref{req}}), take the following form:
\be \label{y}
\frac{dy}{d N}=-\frac{4\left[y+\eta
(Y_{1}+Y_{2})-\eta^{2}(Y_{3}-Y_{4})\right]+\frac{\Gamma\, \zeta_{0}}{H_{0}\,
         \zeta}\left[(y+1)+\eta
(Y_{1}+2Y_{2})-\eta^{2}(Y_{3}-2Y_{4})\right]}{(y+2)+\eta
(Y_{1}+3Y_{2}-2Y_1^2)-\eta^{2}(Y_{3}-3Y_{4}+4 (Y_1-4)Y_1Y_2^3)},
\ee
\be \label{x}
\frac{dx}{d N}= \frac{\Gamma\, \zeta_{0}}{H_{0}\, \zeta}e^{y}
   \left[(y+1)+\eta(Y_{1}+2Y_{2})-\eta^{2}(Y_{3}-2Y_{4})\right]-3x,
\ee
\be
\label{rr} \frac{dr}{d N}= -4r,
\ee
where
\be \label{ze}
\zeta=\sqrt{e^{y}[(y+1)+\eta(Y_{1}+2Y_{2})-\eta^{2}(Y_{3}-2Y_{4})]+x+r},
\ee
$\zeta_{0}=\zeta(z=0)$, and $H_0=H(z=0)$.
Once the parameters $\Gamma$ and $\delta$,
as well as the initial conditions, are specified,
the solution of this set of equations follows immediately.
As our calculation will show,
for the YMC to be a sensible model of dynamic DE,
the order of magnitude of
the decay rate $\Gamma$ should be less than,
or at most of order of the expansion rate, i.e., $\Gamma \le H_0$.
To be specific, we take the decay rate $\Gamma=0.31H_{0}$ and the parameter $\delta=4$.
The initial conditions for Eqs.(\ref{y}) through (\ref{rr})
are chosen at a very high redshift $z_i= 10^{10}$ during the Big Bang nucleosynthesis (BBN) era.
To ensure the equality of radiation-matter occurring at a redshift $z= 3454$ \cite{spergel},
the initial radiation and matter are taken as
\be \label{xi} x_i=1.22 \times 10^{29}, \,\,\,\,
r_i= 3.52\times 10^{35}.
\ee
Besides, to ensure the BBN occurs as usual \cite{walker},
the initial YMC fraction should be $\sim 10\% $ or less \cite{Xia};
for concreteness we take the upper limit
\be \label{yi} y_i \leq 74, \,\,\,\,\, {\rm i.e.}, \,\,\,\,\,
\frac{\rho_{yi}}{\rho_{ri}}\leq 3 \times 10^{-2}  .
\ee

\begin{figure}
\centerline{\includegraphics[width=10cm]{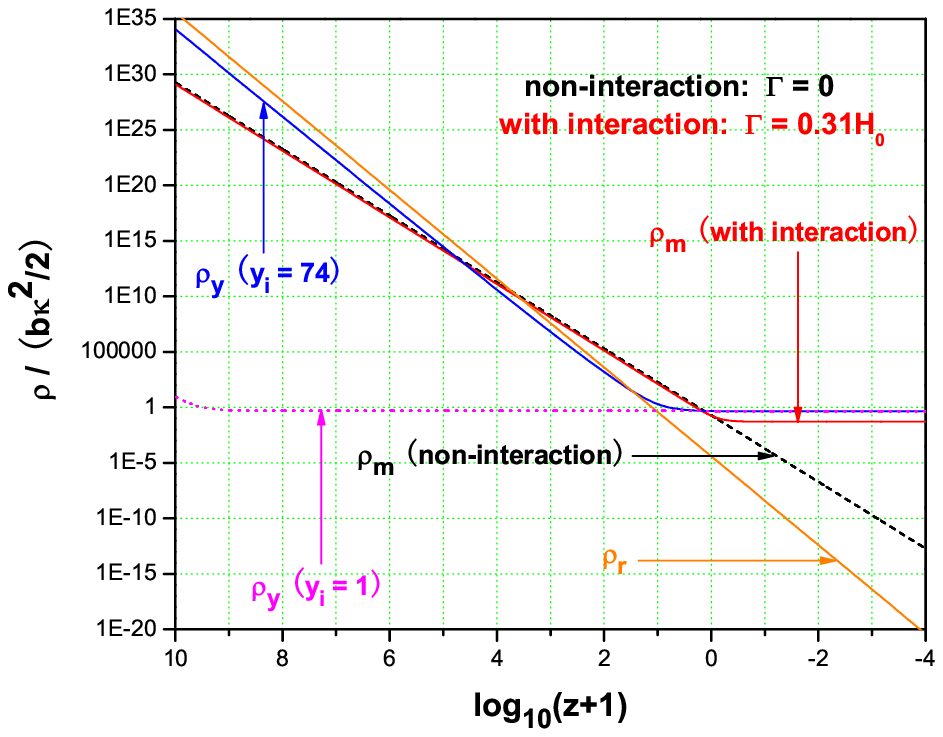}} \caption{
\label{fig1} Dynamical evolution of $\rho_y$, $\rho_m$, and $\rho_r$
in 3-loop YMC model without and with interaction.}
\end{figure}

In Fig.\ref{fig1},
We plot the dynamical evolution of $\rho_y$, $\rho_m$, and $\rho_r$
in 3-loop YMC model without and with interaction.
For a whole range of initial $ y_i=(1, 74)$,
which corresponds to the initial energy fraction of YMC
$\frac{\rho_{yi}}{\rho_{ri}}\simeq(2.5\times 10^{-35}, 3 \times 10^{-2})$
ranging $\sim 33$ orders in magnitude,
YMC always has a desired tracking solutions,
i.e., during the radiation era the YMC follows the radiation as $\rho_y\propto \rho_r \propto a(t)^{-4}$,
then during the matter era it follows the matter approximately as $\rho_y \propto \rho_m \propto a(t)^{-3}$.
Rather later around $z\simeq0.5$ it becomes dominant,
and then it levels off and becomes asymptotically a constant for $z\leq 0$.
The existence of a tracking solution can be analytically proved, also.
For any DE model, the energy density and the pressure of DE can be written as
\be  \label{rhop}
\rho_{y}=E_{k}+V(y),~~~p_{y}=E_{k}-V(y),
\ee
where $E_{k}$ denotes the kinetic energy,
and $V(y)$ denotes the potential energy.
For our model, one can easily obtain $V$ from Eqs.(\ref{energy}) and (\ref{pressure}).
Following the Ref.\cite{Steinhardt},
we introduce a key function
\be  \label{key}
\Delta \equiv \frac{V'' V}{(V')^{2}},
\ee
whose properties determine whether tracking solutions exist.
The prime means derivative with respect to $y$.
Steinhardt et.al. \cite{Steinhardt} had demonstrated that,
the tracking solutions exist if:
(1) $\Delta$ is nearly constant,
i.e., $\mid \Delta^{-1}\frac{d(\Delta-1)}{Hdt}\mid \approx \mid \frac{\Delta'}{\Delta(V'/V)} \mid \ll 1$;
and (2) $\Delta > 1$ for $w < w_{B}$ or $\Delta < 1$ for $w_{B} < w < (1/2)(1+w_{B})$.
(Here $w_{B}$ denotes the EOS of background component)
Our calculation shows that for the range $y>1$ in YMC model,
$\mid \frac{\Delta'}{\Delta(V'/V)} \mid \ll 1$ is always satisfied;
besides, $w < w_{B}$ and $\Delta > 1$ also hold true in our model.
Therefore, there is a tracking solution exists in our model.
For a smaller initial value $\rho_{yi}$,
$\rho_y(t)$ still tracks $\rho_r(t)$,
but for a shorter period correspondingly,
and then approaches to the same constant.
When the initial value  $ y_i$ is sufficient small,
the YMC dark energy is effectively similar to the cosmological constant $\Lambda$.
For the non-interaction case, the matter component retains $\rho_m \propto a(t)^{-3}$ and always decays with time $t$.
For the case of YMC decaying into matter,
the matter density $\rho_{m}$ deviates from $\propto a(t)^{-3}$ around $z\sim0$ and becomes a constant at last.
For the decay rate $\Gamma=0.31H_{0}$,
the dynamical equation for $(x,y)$ at $t\rightarrow \infty$ has a fixed point $(x_f,y_f)=(0.052,-0.887)$,
regardless of initial conditions.
Let we study the stability of this fixed point analytically.
Base on Eqs.(\ref{x}) and (\ref{y}),
one can obtain,
\be  \label{xy}
\frac{d x}{d N}=f(x,y),~~~\frac{d y}{d N}=g(x,y),
\ee
where $f(x,y)$ and $g(x,y)$ are the terms on the right hand side of  Eqs. (\ref{x}) and (\ref{y}), respectively.
By a standard procedure,
expanding  $x=x_f+\varepsilon$ and $y=y_f+\eta$
($\varepsilon$ and $\eta$ are small perturbations around the fixed point),
and keeping up to the first order of small perturbations,
Eq.(\ref{xy}) is reduced to
\be  \label{M}
\frac{d}{d N}
\left(
 \begin{array}{c}
 \varepsilon\\
 \eta
  \end{array}
 \right)
 = M
 \left(
 \begin{array}{c}
 \varepsilon\\
 \eta
  \end{array}
 \right),
\ee
where $M$ is a $2\times 2$ matrix, whose elements are
$M_{11}=\frac{\partial f(x_{f},y_{f})}{\partial x}$,
$M_{12}=\frac{\partial f(x_{f},y_{f})}{\partial y}$,
$M_{21}=\frac{\partial g(x_{f},y_{f})}{\partial x}$,
and $M_{22}=\frac{\partial g(x_{f},y_{f})}{\partial y}$, respectively.
The general solution
for the linear perturbations is of the form
\be
\varepsilon=C_1e^{\mu_1 N}+C_2e^{\mu_2 N},~~~\eta=C_3e^{\mu_1 N}+C_4e^{\mu_2 N},
\ee
where $C_1$, $C_2$, $C_3$, and $C_4$ are constants,
$\mu_1$ and $\mu_2$ are the eigenvalues of matrix $M$.
As long as $\mu_1$ and $\mu_2$ are both negative,
the fixed point $(x_f, y_f)$ is stable,
and the solution is an attractor.
By Calculating, we find the matrix
\be
M= \left(
\begin{array}{cc}
-3.16068 &  0.18234 \nonumber\\
 0.20061 &  -2.55983
\end{array}
\right),
\ee
and its two eigenvalues are
$\mu_1=-2.50411$ and $\mu_2=-3.21640$, respectively, both negative.
Thus the fixed point of this model is stable against perturbation,
and the solution is an attractor.
Therefore, the Big Rip would not happen in our model.
For an initial YMC DE ranging $\sim 33$ orders in magnitude,
and for both cases without and with interaction,
the fractional densities $\Omega_{y 0}=0.73$ and $\Omega_{m 0}=0.27$ are always achieved at $z=0$.
In addition, from the theoretical point of view \cite{Steinhardt},
for any DE model with an attractor-like behavior,
as long as $w\leq w_{B}$ is satisfied,
the coincidence problem can be naturally solved.
Both these two conditions are satisfied in our model.
Therefore, the coincidence problem is naturally solved in 3-loop YMC model.

\begin{figure}
\centerline{ \includegraphics[width=10cm]{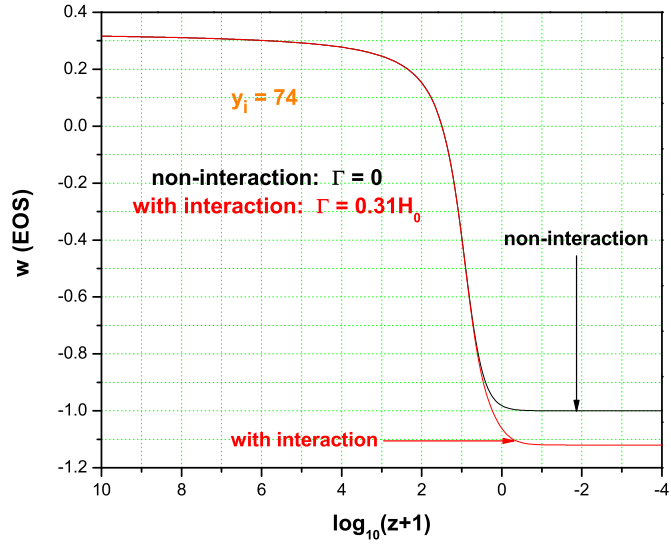}}
\caption{\label{fig2} Evolution of EOS $w$ in 3-loop YMC model
without and with interaction.}
\end{figure}

In Fig.\ref{fig2},
we plot the evolution of EOS $w(z)$ in 3-loop YMC model without and with interaction.
At the early stage of Universe (high energies limit),
$w(z)$ approaches to that of radiation, i.e., $w\rightarrow1/3$.
With the expansion of Universe and the decreasing of energy scale,
$w$ smoothly decreases, and the YMC component transits from radiation to matter, and to DE.
For the non-interaction case, $w$ does not cross, only asymptotically approaches to $-1$.
For the interaction case, $w$ can smoothly cross $w = -1$,
as indicated by the preliminary observational data \cite{HST,SNLS,ESSENCE}.
Adopting the initial $y_i=74$,
the EOS of YMC will cross $-1$ around $z \simeq0.6$.
For a smaller $y_i$ the crossing occurs earlier.
For example, taking  $y_i= 72$ in our model will make $w$ crossing $-1$ occur around $z \simeq 1.5$,
as suggested in Ref.\cite{Hao1}.
Moreover, treating $\Gamma$ as a parameter,
we find that a constant interaction $\Gamma$ corresponds to a constant present EOS $w_{0}$,
and a larger $\Gamma$ yields a smaller $w_{0}$.
For instance,
$\Gamma=0.31H_{0}\rightarrow w_{0}=-1.05$;
$\Gamma=0.67H_{0}\rightarrow w_{0}=-1.15$;
and  $\Gamma=0.82H_{0}\rightarrow w_{0}=-1.21$.

In comparison with the 2-loop YMC model,
the transition to the DE-dominant era
occurs at $z\simeq0.5$ in the 3-loop  model,
later than $z\simeq0.6$ predicted by the 2-loop model \cite{Xia}.
Besides, 3-loop model yields an  EOS $w_{0}=-1.06$,
which is closer to the observational constraints on $w_{0}$ \cite{HST}
than that of the 2-loop model \cite{Xia}.
In addition, the 3-loop model can also predict a larger age of the Universe than the 2-loop model \cite{Wang}.
Notice that the YMC is subdominant during the early stages,
so the nucleosynthesis and the recombination occur as in the standard Big Bang cosmology.
Besides, since the DE becomes dominant at very late era,
the matter era is also long enough for the structure formation.
It should be mentioned that the scale $\kappa$ can be fixed
by requiring $\rho_{y}$ in Eq.({\ref{energy}})
be equal to the dark energy density $\sim 0.73 \rho_{c}$,
where $\rho_{c}$ is the critical density,
yielding $\kappa^{1/2}\simeq 7.5 h_{0}^{1/2}\times10^{-3} eV$ ($h_{0}$ is the Hubble parameter).
At the moment we do not have an answer to the question why $\kappa$ is so small,
so the fine-tuning problem is still present in our model.
As has been shown \cite{zhang0},
in the case of the YMC decaying into matter and radiation,
both $\rho_m$ and $\rho_y$ will asymptotically approach to their respective constant values,
i.e., the future of the universe is a steady state,
quite similar to that of the Steady State model \cite{bondi}.
Therefore, in a sense, our model bridges  between the Big Bang and the Steady State model.

\

\begin{center}
{\em\Large 3. The diagnosis of $H$, $q$,
$r-s$, and $w-w^\prime$ in the Model}
\end{center}

\begin{figure}
\centerline{ \includegraphics[width=10cm]{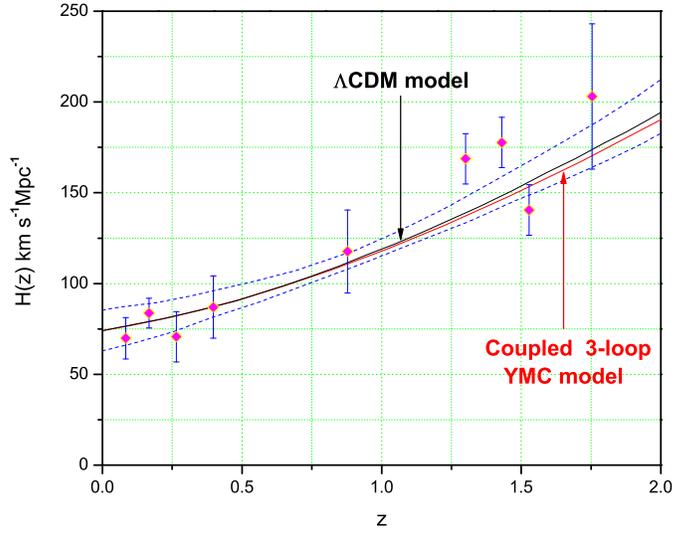}}
\caption{\label{fig3} Comparison of the observed $H(z)$ in square dots
with the predictions by $\Lambda$CDM model \cite{Simon} and by coupled 3-loop YMC model.}
\end{figure}

In this section the decay rate is taken to be $\Gamma=0.31 H_0$ as before.
First, let us discuss the Hubble parameter $H\equiv \dot{a}/a$.
The expansion of Universe is determined by the Friedmann equations
\be \label{Fried1}
H^{2}=\frac{8\pi G}{3}\rho,
\ee
\be
\label{Fried2} \frac{\ddot{a}}{a}= -\frac{4\pi G}{3}(\rho+3p),
\ee
where the total energy density $\rho=\rho_y+\rho_m+\rho_r$,
and the total pressure $p=p_{y}+p_{r}$.
The dynamical evolution of $\rho_y$, $\rho_m$, and $\rho_r$ have already been given in the previous section,
so the Hubble parameter $H$ can be easily obtained.
In Fig.\ref{fig3} we compare the observed expansion rate $H(z)$ \cite{Simon}
with that predicted by $\Lambda$CDM model \cite{Simon,Hao1} and by coupled 3-loop YMC model.
The area surrounded by two dashed lines
shows the $68\%$ confidence interval \cite{spergel}.
It is seen that the coupled YMC model is quite close to the $\Lambda$CDM model in the range $z\leq1$,
and both two models approximately agree with the observations.
The observed dip of $H(z)$ around $z\sim 1.5$ \cite{Simon} is difficult for both models,
but $H(z)$ in our model is slightly lower than that in $\Lambda$CDM model and is closer to the dip.
For the non-interaction case in the 3-loop YMC model,
$H(z)$ is more close to the $\Lambda$CDM model.
Since its evolution trajectory is almost overlaps with that of $\Lambda$CDM model,
we do not plot its curve here.

\begin{figure}
\centerline{ \includegraphics[width=10cm]{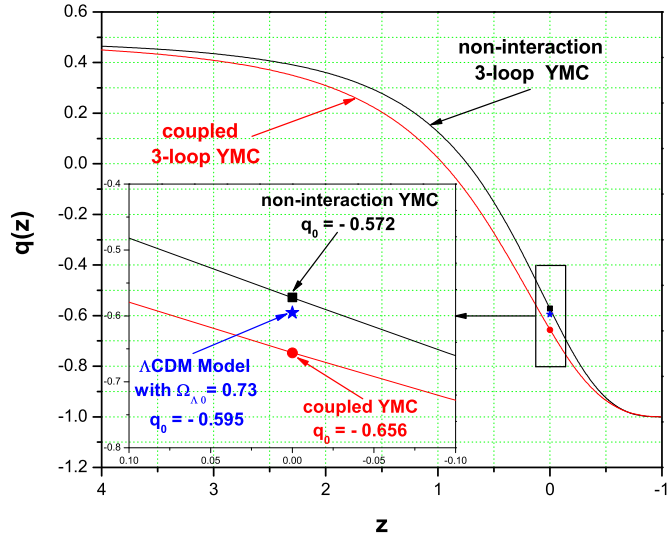}}
\caption{\label{fig4} Evolution of deceleration parameter $q(z)$
in 3-loop YMC model without and with interaction.}
\end{figure}

Next, we turn to the deceleration parameter $q(z)$, which is given by
\be  \label{q}
q\equiv-\frac{\ddot{a}}{aH^2}=\frac{1}{2}(1+3\Omega_y w+\Omega_r),
\ee
where $\Omega_y=\rho_y/\rho$ and $\Omega_r=\rho_r/\rho$.
In deriving this expression,
the Eqs.(\ref{w}), (\ref{Fried1}), and (\ref{Fried2}) are used.
In Fig.\ref{fig4} we plot $q(z)$ in 3-loop YMC model without and with interaction.
In both cases,
starting from a positive value during the matter era,
$q(z)$ decreases with the expansion of Universe,
turns into negative around $z\sim 1$,
and approaches to an asymptotic value $q=-1$ in future.
The current value is $q_{0}=-0.572$ for the non-interacting case,
which is denoted by a square dot;
and is $q_{0}=-0.656$ for the interacting case,
which is denoted by a round dot.
In comparison,
the $\Lambda$CDM model with $\Omega_{\Lambda }=0.73$ has $q_{0}=-0.595$, denoted by a star symbol.
so the non-interaction YMC model is closer to the $\Lambda$CDM model than the coupled YMC.

\begin{figure}
\centerline{ \includegraphics[width=10cm]{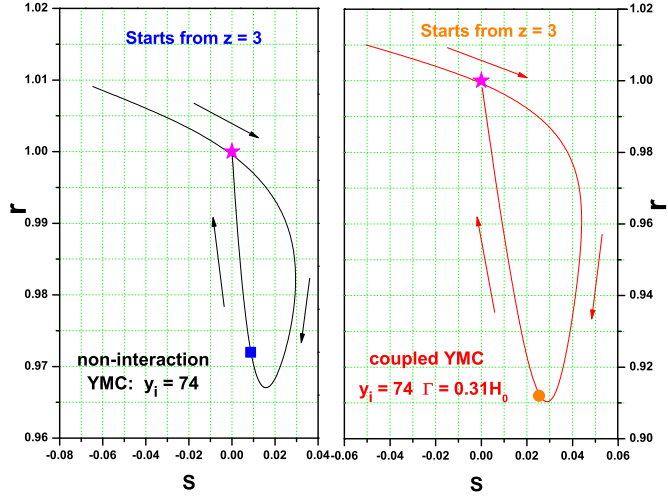}}
\caption{\label{fig5} Evolution trajectories of statefinder in $r-s$ plane
for 3-loop YMC model without and with interaction.}
\end{figure}

Now, we study the statefinder diagnosis defined as \cite{Sahni}
\be \label{statefinder}
r\equiv\frac{\stackrel{\dots}{a}}{aH^3},~~~~~
s\equiv\frac{r-1}{3(q-1/2)},
\ee
Taking time derivative of Eq.(\ref{Fried2})
and making use of Eqs.(\ref{ymeq}), (\ref{meq}), and (\ref{req}),
one obtains
\be\label{r}
r=1+\frac{9}{2}\Omega_y  w(1+w)
  -\frac{3}{2}\Omega_y w^\prime +2\Omega_r
  +  \frac{3\Gamma}{2H}\Omega_y w
\ee
\be\label{s}
s=\frac{3\Omega_y w (1+w)-\Omega_y w^\prime
    +\frac{4}{3}\Omega_r
    +\frac{\Gamma}{H}\Omega_y w}
    {3\Omega_y w +\Omega_r},
\ee
where
\be\label{wy}
w^\prime \equiv \frac{dw}{dN}=\frac{dw}{dy}\frac{dy}{dN}
\ee
can be calculated by Taking variable $y$ derivative of Eq.(\ref{w}).
The expressions of $r$ and $s$ in
Eqs.(\ref{r}) and (\ref{s}) hold actually for a generic DE model.
Since different cosmological models exhibit
qualitatively different trajectories of evolution in the $r-s$ plane,
the statefinder is a useful tool to distinguish cosmological models \cite{Sahni}.
In Fig.\ref{fig5} we plot the evolution trajectories of statefinder in $r-s$ plane
for 3-loop YMC model without and with interaction,
starting from the redshift $z=3$.
The arrows alone the curves indicate the direction of evolution.
The overall profile of statefinder diagnosis predicted by these two model are quite similar.
As the Universe expands,
$s$ increases to a maximum and $r$ decreases to a minimum;
after that, the trajectories turn a corner and approach to a final fixed point $(r,s)=(1,0)$.
This fixed point is same as that predicted by $\Lambda$CDM model \cite{Sahni},
denoted by a star symbol on the plot.
The current value of statefinder is $(r,s)=(0.972, 8.660\times10^{-3})$ for the non-interacting case,
and is $(r,s)=(0.912, 2.528\times10^{-2})$ for the interacting case.
Base on these calculated values,
one sees that the non-interacting 3-loop YMC model is closer to the $\Lambda$CDM model;
and the introduction of interaction between matter and DE cause a deviation from the $\Lambda$CDM model.
We also find that a larger interaction $\Gamma$ yields a larger deviation.
For instance, $\Gamma = 0.67H_{0}$ yields $(r,s)=(0.851, 3.898\times10^{-2})$ at $z=0$,
which is further away from the value of $\Lambda$CDM.
In comparison,
the quietessence model \cite{Alam} gives the current values $(r,s)=(0.4,0.3)$,
the Chaplygin gas model \cite{Gorini} gives $(r,s)=(1.95,-0.3)$,
and the agegraphic model \cite{Hao2} with parameter $n=2.0$ give $(r,s)\simeq(-0.2,0.5)$.
All the current values of $(r,s)$ predicted by these three types of DE models
are far away from the $(1,0)$ given by $\Lambda$CDM model.
The holographic model \cite{JFZ}
without interaction gives $(r,s)\simeq(0.94,0.01)$,
when an interaction is included, it gives $(r,s)\simeq(0.75,0.09)$,
deviating away considerably from the $\Lambda$CDM model again.
Therefore, our model is much closer to $\Lambda$CDM than other dynamics DE models.

\begin{figure}
\centerline{ \includegraphics[width=10cm]{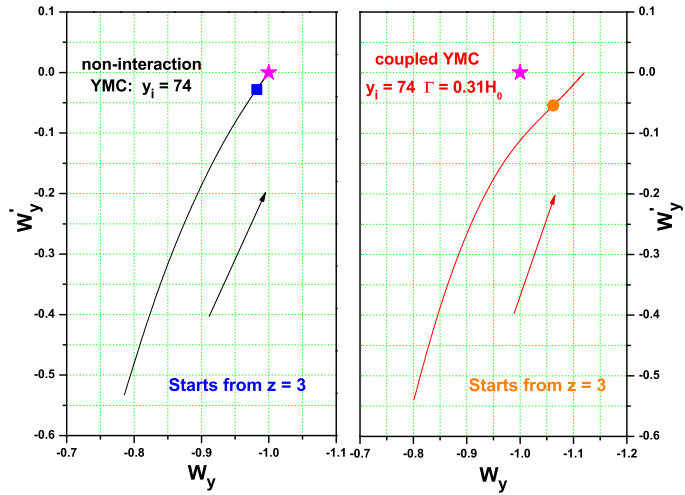}}
\caption{\label{fig6} Evolution trajectories of $w$-$w^{'}$
for 3-loop YMC model without and with interaction.}
\end{figure}

Finally, we investigate the $w-w^\prime$ diagnosis,
defined in the Eqs.(\ref{w}) and (\ref{wy}).
In Fig.\ref{fig6} we plot the evolution trajectories of $w-w^\prime$
for 3-loop YMC model without and with interaction,
starting from the redshift $z=3$.
The arrows alone the curves denote the direction of evolution.
For both models,
with the expansion of the Universe,
$w$ decreases and $w^\prime$ increases,
and the $w-w^\prime$ diagnosis approaches to a fixed point asymptotically.
For the non-interacting case, the current value is $(w,w') = (-0.982,-2.778\times10^{-2})$,
and the fixed point is $(-1,0)$,
which is same as that of $\Lambda$CDM model, denoted by a star symbol on the plot.
Therefore, the non-coupling YMC model can not cross the phantom divide $w=-1$.
For the interacting case, the situation is different:
$\Gamma= 0.31H_0$ yields
a current value $(w,w') = (-1.063,-5.430\times10^{-2})$,
and approaches to a fixed dot $(-1.12, 0)$ asymptotically.
Therefore,
the interaction between matter and DE
causes a deviation from the $\Lambda$CDM,
and give an EOF crossing the phantom divide $w=-1$.

\

\begin{center}
{\em\Large 4. Cosmological Constraints From SNIa, CMB, and LSS}
\end{center}

In the following,
by using the maximum likelihood method,
we will perform the best fit analysis on our 3-loop YMC model
with the latest observational data of SNIa, CMB and LSS.
First, we derive the constraints on the model from SNIa.
Recently, the up-to-date gold sample of 182 SNIa data was compiled by Riess et al. \cite{Riess2}.
It provides the apparent magnitude $m(z)$ of supernovae,
which is related to the luminosity distance $d_{L}(z)$ of supernovae through
\be\label{m}
m(z)=M+5\log d_L(z)+25,
\ee
where $M$ is the absolute magnitude,
that can generally be considered to be the same for SNIa.
In a flat universe the luminosity distance satisfies
\be\label{d}
d_L=H_0^{-1}(1+z)\int_0^z\frac{dz'}{E(z')},
\ee
where $E(z)\equiv H(z)/H_{0}$ and the Hubble scale $H^{-1}_0=2997.9h^{-1}Mpc$.
The data points of the latest 182 SNIa Gold dataset compiled in \cite{Riess2}
are given in terms of the distance modulus
\be\label{mu1}
\mu_{obs}(z_i)\equiv m_{obs}(z_i)-M.
\ee
On the other hand, the theoretical distance modulus is defined as
\be\label{mu2}
\mu_{th}(z_i)\equiv m_{th}(z_i)-M=5\log_{10}d_L(z_i)+25.
\ee
The theoretical model parameters are determined by minimizing
\be\label{chiSN}
\chi^2_{SN}({\bf p})=\sum\limits_{i=1}^{182}
\frac{\left[\mu_{obs}(z_i)-\mu_{th}(z_i)\right]^2}{\sigma^2(z_i)},
\ee
where $\sigma$ is the corresponding $1\sigma$ error,
and ${\bf p}$ denotes the model parameters.
In this work,
we will determine the best fit values of corresponding model parameters
(including the present fractional matter density $\Omega_{m0}$,
the Hubble constant $h$, and the decay rate $\Gamma$) in 3-loop YMC model.

Next,
we also consider the constraints from CMB \cite{spergel} and LSS \cite{bahcall} observations.
For the CMB data, we use the CMB shift parameter $R$,
which is perhaps the most model-independent parameter that can be extracted from CMB data.
The CMB shift parameter is given by \cite{Tegmark}
\be\label{R}
R\equiv\Omega_{m0}^{1/2}\int_0^{z_{rec}}
\frac{dz'}{E(z')},
\ee
where the redshift of recombination $z_{rec}=1090$, which is given by WMAP5 \cite{WMAP5}.
The shift parameter $R$ relates the angular diameter distance to the last scattering surface,
the comoving size of the sound horizon at $z_{rec}$ and the angular scale of the first acoustic peak
in CMB power spectrum of temperature fluctuations \cite{Tegmark,ywang}.
The measured value of $R$ has been updated to be $R_{obs}=1.710\pm 0.019$ from WMAP5 \cite{WMAP5}.
On the other hand, for the LSS data,
we use the parameter $A$ from the measurement of BAO peak
in the distribution of SDSS luminous red galaxies,
which is given by \cite{Eisenstein}
\be\label{A}
A\equiv\Omega_{m0}^{1/2}E(z_b)^{-1/3}\left[\frac{1}{z_b}
\int_0^{z_b}\frac{dz'}{E(z')}\right]^{2/3},
\ee
where $z_b=0.35$. The SDSS BAO measurement \cite{Eisenstein} gives
$A_{obs}=0.469\,(n_s/0.98)^{-0.35}\pm 0.017$.
here the scalar spectral index $n_s$ is taken to be $0.96$ from the 5-year WMAP data \cite{WMAP5}.
Since both parameters, $R$ and $A$,
are independent of Hubble constant $H_{0}$
and can be easily obtained from CMB and LSS observations,
they provide robust constraints on DE models as useful complements to the SNIa data.
We perform a joint analysis of latest 182 SNIa Gold dataset, the shift parameter $R$ from CMB
and the BAO peak measurement $A$ from LSS to constrain the 3-loop YMC model.
The total $\chi^2$ is given by
\be\label{chitotal}
\chi^2=\chi^2_{SN}+\chi^2_{CMB}+\chi^2_{LSS},
\ee
where $\chi^2_{SN}$ is given by Eq.(\ref{chiSN}),
and the latter two terms are defined as
\be\label{chiCMB}
\chi^2_{CMB}=\frac{(R-R_{obs})^2}{\sigma_R^2},
\ee
and
\be\label{chiLSS}
\chi^2_{LSS}=\frac{(A-A_{obs})^2}{\sigma_A^2}.
\ee
The corresponding $1\sigma$ errors are $\sigma_R=0.03$ and $\sigma_A=0.017$, respectively.
As usual, assuming the measurement errors are Gaussian, the likelihood function
\be\label{likelihood}
{\cal{L } } \propto e^{-\chi^2/2}.
\ee
The model parameters that yielding a minimal $\chi^{2}$ and a maximal ${\cal{L } }$
will be favored by the observations.
The results of our analysis are presented as the following.

\begin{figure}
\centerline{ \includegraphics[width=10cm]{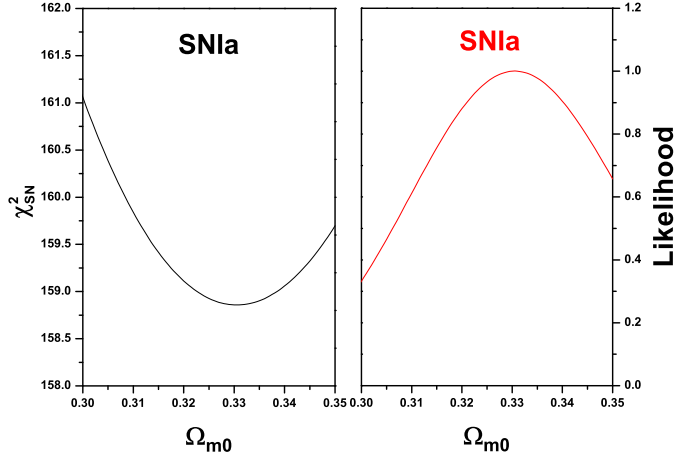}}
\caption{\label{fig7} The $\chi_{SN}^{2}$ and the corresponding likelihood ${\cal{L } }$ of 3-loop YMC,
where the best-fit parameter $h=0.626$ is adopted.
These results are obtained from the latest 182 SNIa Gold dataset.}
\end{figure}

First for the non-interacting 3-loop YMC model.
Utilizing the latest 182 SNIa Gold dataset alone,
we plot the $\chi_{SN}^{2}$ and the corresponding likelihood ${\cal{L } }$ of 3-loop YMC in Fig.\ref{fig7}.
It is found that the best-fit model parameters are $\Omega_{m0}=0.331$ and $h=0.626$,
giving a minimal $\chi_{SN}^{2}=158.858$.
Besides, for $68\%$ confidence level \cite{Nesseris},
the range of parameters are determined as
$\Omega_{m0}=0.331^{+0.020}_{-0.021}$, $\Omega_{y0}=0.669^{+0.021}_{-0.020}$ and $h=0.626\pm0.004$.
In comparison, we also fit the $\Lambda$CDM model to the same SNIa data
and find that the minimal $\chi_{SN,\Lambda}^{2}=158.750$
for the best-fit parameter $\Omega_{m0}^{\Lambda}=0.344$ and $h=0.626$.
So fitting to the 182 SNIa Gold dataset alone,
the $\Lambda$CDM model is slightly better than the 3-loop YMC model.

\begin{figure}
\centerline{ \includegraphics[width=10cm]{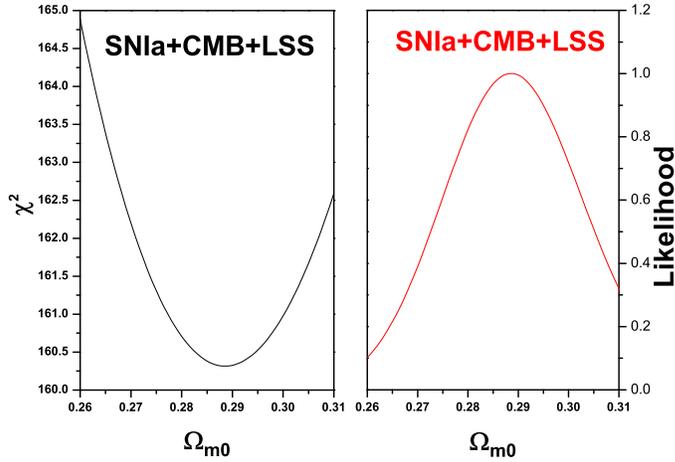}}
\caption{\label{fig8} The $\chi^{2}$ and the corresponding likelihood ${\cal{L } }$ of 3-loop YMC,
where the best-fit parameter $h=0.634$ is adopted.
These results are obtained from the combined SNIa, CMB and LSS data.}
\end{figure}

Moreover, utilizing the combination of SNIa, CMB and LSS data,
we plot the $\chi^{2}$ and the corresponding likelihood ${\cal{L } }$ of 3-loop YMC in Fig.\ref{fig8}.
It is found that the best-fit model parameters are $\Omega_{m0}=0.289$ and $h=0.634$,
giving a minimal $\chi_{min}^{2}=160.317$.
Besides, for $68\%$ confidence level \cite{Nesseris},
the range of parameters are determined as
$\Omega_{m0}=0.289^{+0.013}_{-0.014}$, $\Omega_{y0}=0.711^{+0.014}_{-0.013}$ and $h=0.634\pm0.004$.
As a comparison, we also fit the $\Lambda$CDM model to the same combined data.
It is found that the minimal $\chi_{min,\Lambda}^{2}=162.460$
for the best-fit parameter $\Omega_{m0}^{\Lambda}=0.283$ and $h=0.638$.
Therefore, on fitting to the combination of SNIa, CMB and LSS data,
the non-interacting 3-loop YMC model is better than the $\Lambda$CDM.
This makes the 3-loop YMC model more attractive.

Finally, we turn to the coupled YMC model
and constrain the decay rate $\Gamma$ as the last model parameter.
In Table 1,
we list the minimal $\chi_{SN,min}^{2}$ and $\chi_{min}^{2}$ for 3-loop YMC model with various $\Gamma$.
For the latest SNIa data alone,
the non-interaction YMC model is only slightly better than the coupled YMC.
However, for the combination of SNIa,  CMB and LSS data,
the non-interaction YMC model is much more favored than the coupled YMC.
This fact can be explained as follows:
It is well known that most current observations,
including SNIa, CMB and LSS,
all favor the $\Lambda$CDM model.
Therefore, to fit these observations well,
a DE model should not deviate too far away from the $\Lambda$CDM model.
Since the introduction of interaction between matter and DE
will cause a deviation from the $\Lambda$CDM model,
and a larger interaction $\Gamma$ yields a larger deviation,
the coupled YMC model with a large $\Gamma$ will not be favored by the observations.
That is to say,
to consistent with the current observations,
the DE-matter interaction (if it exists) should be very small.

\begin{table}
\caption{The minimal $\chi_{SN,min}^{2}$ and $\chi_{min}^{2}$ for 3-loop YMC model with various $\Gamma$.}
\begin{center}
\label{Gamma}
\begin{tabular}{|c|c|c|c|}
  \hline
  $\Gamma$& 0 & $0.12H_{0}$ & $0.18H_{0}$ \\
  \hline
  $\chi_{SN,min}^{2}$ & 158.858 & 158.877 & 158.950 \\
  \hline
  $\chi_{min}^{2}$ & 160.317 & 174.498 & 192.018 \\
  \hline
\end{tabular}
\end{center}
\end{table}

\

\begin{center}
{\em\Large 5. Summary}
\end{center}

In this work, we extend our previously proposed YMC DE model to the 3-loop quantum corrections.
This model can naturally solve the coincidence problem,
and it can give, in the interaction form, an EOF crossing the phantom divide $w=-1$.
Next, we study the Hubble parameter $H$, the deceleration parameter $q$,
the statefinder diagnosis $(r,s)$, and the $w-w^\prime$ diagnosis of 3-loop YMC model
for both cases without and with interaction,
and compare our results with other DE models.
It is found that the 3-loop YMC model is much closer to the $\Lambda$CDM model than other dynamics DE models;
and the introduction of the matter-DE interaction will make the YMC model deviating from the $\Lambda$CDM model.
Finally, by using the observational data of SNIa,
the shift parameter from CMB,
and the BAO peak from LSS,
we give the cosmological constraints on 3-loop YMC model.
Utilizing the latest SNIa data alone,
the best-fit model parameters of 3-loop YMC are
$\Omega_{m0}=0.331^{+0.020}_{-0.021}$, $\Omega_{y0}=0.669^{+0.021}_{-0.020}$ and $h=0.626\pm0.004$
(with $1\sigma$ uncertainty);
and combining SNIa, CMB and LSS data,
the best-fit model parameters are
$\Omega_{m0}=0.289^{+0.013}_{-0.014}$, $\Omega_{y0}=0.711^{+0.014}_{-0.013}$ and $h=0.634\pm0.004$
(with $1\sigma$ uncertainty).
To fit the latest SNIa data alone,
the $\Lambda$CDM model is slightly better than the 3-loop YMC model;
but in fitting of the combination of SNIa, CMB and LSS data,
the 3-loop YMC model performs better than the $\Lambda$CDM model.
This makes the 3-loop YMC model more attractive.
In addition, the maximum likelihood analysis also shows that
the interaction between matter and DE should be small.

There are also other observations that would be helpful to constraint the DE models,
such as the Chandra X-Ray observation \cite{Allen},
the lookback time data \cite{Capozziello},
the Gamma-Ray Bursts \cite{Dai} and so on.
In addition, it is also very interesting to constrain the YMC model
by using the global fitting to the full CMB and LSS data via Markov Chain Monte Carlo analysis.
These issues deserve further investigations in the future.

\

ACKNOWLEDGMENT: We are grateful to the Referee for valuable suggestions.
We also thank Dr. W. Zhao for helpful discussions.
Y.Zhang's research work is supported by the CNSF No.10773009, SRFDP, and CAS.


\newpage

\end{document}